\documentclass[runningheads]{llncs}
\usepackage{newtxtext}       % 
\usepackage[varvw]{newtxmath}       % selects Times Roman as basic font
\usepackage{amsfonts}
\usepackage{amsmath}

\usepackage{amssymb}
\usepackage{appendix}
\renewcommand{\figurename}{Fig.}
\usepackage{color}
\usepackage{bm}
\usepackage{here}
\usepackage{subcaption}
\usepackage{graphicx}
\usepackage{float}
\begin{document}
\title{Load balancing in parallel infinite-server queues with action delay via phase representation}
\titlerunning{Load balancing in parallel infinite-server queues with action delay}
\author{
Kazuma Abe\inst{1}%\orcidID{0000-0002-3916-0904} 
\and
Tuan Phung-Duc\inst{2}%\orcidID{0000-0002-5002-4946}
}
\authorrunning{K. Abe and T. Phung-Duc}
\institute{Graduate School of Science and Technology, University of Tsukuba, Japan \and
Institute of Systems and Information Engineering, University of Tsukuba, Japan
\email{abe.kazuma.su@alumni.tsukuba.ac.jp}\\
\email{tuan@sk.tsukuba.ac.jp}
}
\maketitle             
\begin{abstract}
Spatially distributed systems can use state-dependent routing to allocate users, tasks, or requests to less-loaded nodes. In practice, travel time, communication latency, or actuation can introduce a lag between a routing decision and its execution. We refer to this lag as the action delay. We study state-dependent routing to two parallel infinite-server queues under an action delay. Each arriving job is routed based on the current scaled numbers of jobs in service at the two queues and enters service only after the delay. Representing the Erlang-distributed action delay as a sequence of exponential phases, we obtain a finite-dimensional Markov jump process and ordinary differential equations that explicitly track the jobs in the delay phases. Exploiting the symmetry of the two parallel infinite-server queues, we derive equations governing the imbalance in the numbers of jobs in service and obtain the characteristic equation of the linearized system for an arbitrary number of delay phases. This characteristic equation coincides with that of the corresponding Erlang delayed-information model. Numerical experiments compare the fluid trajectories with stochastic simulations, illustrate how the number of phases and routing sensitivity affect transient oscillations, and show how the local stability boundary varies with the mean action delay.

\end{abstract}

\section{Introduction} \label{sec:intro}

Load balancing is important in spatially distributed service systems where users, jobs, or requests must be assigned to one of several geographically or logically separated service nodes. Such systems arise in mobile edge computing, ride-hailing and delivery platforms, bike-sharing systems, and emergency evacuation or sheltering operations. Routing policies often use real-time congestion or queue-length information to direct jobs to less heavily loaded servers.

A routing decision might not be perfectly aligned with the system state, and the discrepancy takes two distinct forms. The first is an information delay, which occurs when the decision is based on stale congestion information. The second is an action delay. Once a job is assigned, the routing decision takes effect only after signal travel time, communication latency, setup time, or actuation. Thus, information delays affect the input to the routing decision, whereas action delays postpone its execution. Either can weaken the corrective effect of state-dependent routing and induce persistent imbalance or oscillatory behavior across servers.

The effect of delayed information on load balancing has been studied extensively. Mitzenmacher showed that large delays can degrade system performance and induce oscillations in queue length~\cite{mitzenmacher2000useful}. For multinomial-logit routing, Pender et al.\ analyzed parallel infinite-server queues whose routing decisions are based on delayed queue-length information. They identified the condition under which sustained oscillations occur~\cite{pender2017queues}. This framework was later extended to a stochastic setting, where fluid and diffusion approximations were established for systems with delayed information~\cite{pender2020stochastic}. Novitzky and Pender further allowed the information delay to follow a general distribution. They derived the characteristic equation in terms of its Laplace transform, showing that a constant delay is the most destabilizing for a given mean and that an exponentially distributed delay never destabilizes the system~\cite{novitzky2020probabilistic}. 
Beraldi et al.\ showed empirically that stale state information degrades distributed load balancing in fog computing~\cite{beraldi2022impact}, and Tahir et al.\ proposed a learning-based mean-field control policy for load balancing under delayed information~\cite{tahir2022learning}. In both, the lag is attributed to the information used for the decision. 

As in the models of Pender et al.~\cite{pender2017queues} and Novitzky and Pender~\cite{novitzky2020probabilistic}, we model each server as an infinite-server queue. Novitzky et al.\ argued that, at the fluid level, a multi-server queue behaves like an infinite-server queue when the arrival rate is below the service capacity, and that the fluid limit of the Erlang-A queue behaves like an infinite-server queue on either side of the capacity, with the abandonment rate in place of the service rate above it. However, their analysis excludes the boundary case~\cite{novitzky2020limiting}. Whitt~\cite{whitt2021many} discusses how infinite-server models can provide useful approximations for systems with large finite-server service groups. This interpretation motivates our use of the infinite-server model.

The action delay, which is the focus of the present work, has received less attention, even though it is inherent in spatially distributed systems. Novitzky and Pender also discuss travel after committing to a queue as a motivation for delayed information~\cite{novitzky2020probabilistic}. In their model, the routing probabilities depend on averaged past numbers of jobs in service. In our model, routing decisions use the current scaled numbers of jobs in service, and dispatched jobs enter service only after an action delay. 
Closer to our setting, He et al.\ studied randomized routing to remote queues, where a person selects a distant station~\cite{he2025randomized}, and a mean-field model of large-scale task offloading suggests that offloading delay degrades performance~\cite{abe2026mean}. Here, we examine how the delay between assignment and service entry affects the local stability of the symmetric equilibrium under multinomial-logit routing.

Analyzing the action delay requires tracking the jobs while they are delayed. Because the system state evolves during the delay, we explicitly include jobs undergoing the action delay in the system state. Tracking the remaining delay of each dispatched job individually leads to a state space that grows with the number of jobs in the delay phases. To obtain a finite-dimensional Markov model, we assume that the action delay follows an Erlang distribution and track the number of jobs in each delay phase. 

The contributions of this paper are as follows. First, we introduce a system of two parallel infinite-server queues with multinomial-logit routing and action delay. Second, we represent the action delay by an Erlang phase structure and derive a finite-dimensional fluid model that explicitly tracks jobs in the delay phases. This formulation distinguishes jobs already in service from those assigned to a server but not yet in service. Third, using the symmetry of the two-server system, we decompose the fluid dynamics into sum and difference modes. Routing sensitivity acts only through the difference mode. Fourth, we linearize the difference dynamics around the symmetric equilibrium and obtain the characteristic equation for an arbitrary number of phases. Therefore, the phase-state formulation recovers the local stability conditions previously obtained for the Erlang delayed-information model~\cite{novitzky2020probabilistic}. An exponential action delay preserves local stability, whereas Erlang action delays with two or more phases can destabilize the symmetric equilibrium. Finally, we conduct numerical experiments that validate the fluid limit against stochastic simulations and illustrate the imbalance dynamics. 

The remainder of this paper is organized as follows. In Section~\ref{sec:model}, we describe parallel infinite-server queueing systems and define the notation. Section~\ref{sec:fluid} considers the scaled numbers of jobs in service and in the delay phases and derives the ODEs. Section~\ref{sec:linearized_imbalance} investigates differences in the scaled numbers of jobs between the two parallel infinite-server queues, both in service and in each delay phase. We linearize the system and derive the characteristic equation for an arbitrary number of phases. This equation coincides with that of the corresponding Erlang delayed-information model. Section~\ref{sec:numerical} compares the fluid trajectories with stochastic simulations and illustrates how the action delay affects the imbalance dynamics. Finally, in Section~\ref{sec:conclusion}, we conclude the paper and discuss directions for future work.

\section{Model description}
\label{sec:model}

\begin{figure}[H]
    \centering
    \includegraphics[width=0.6\linewidth]{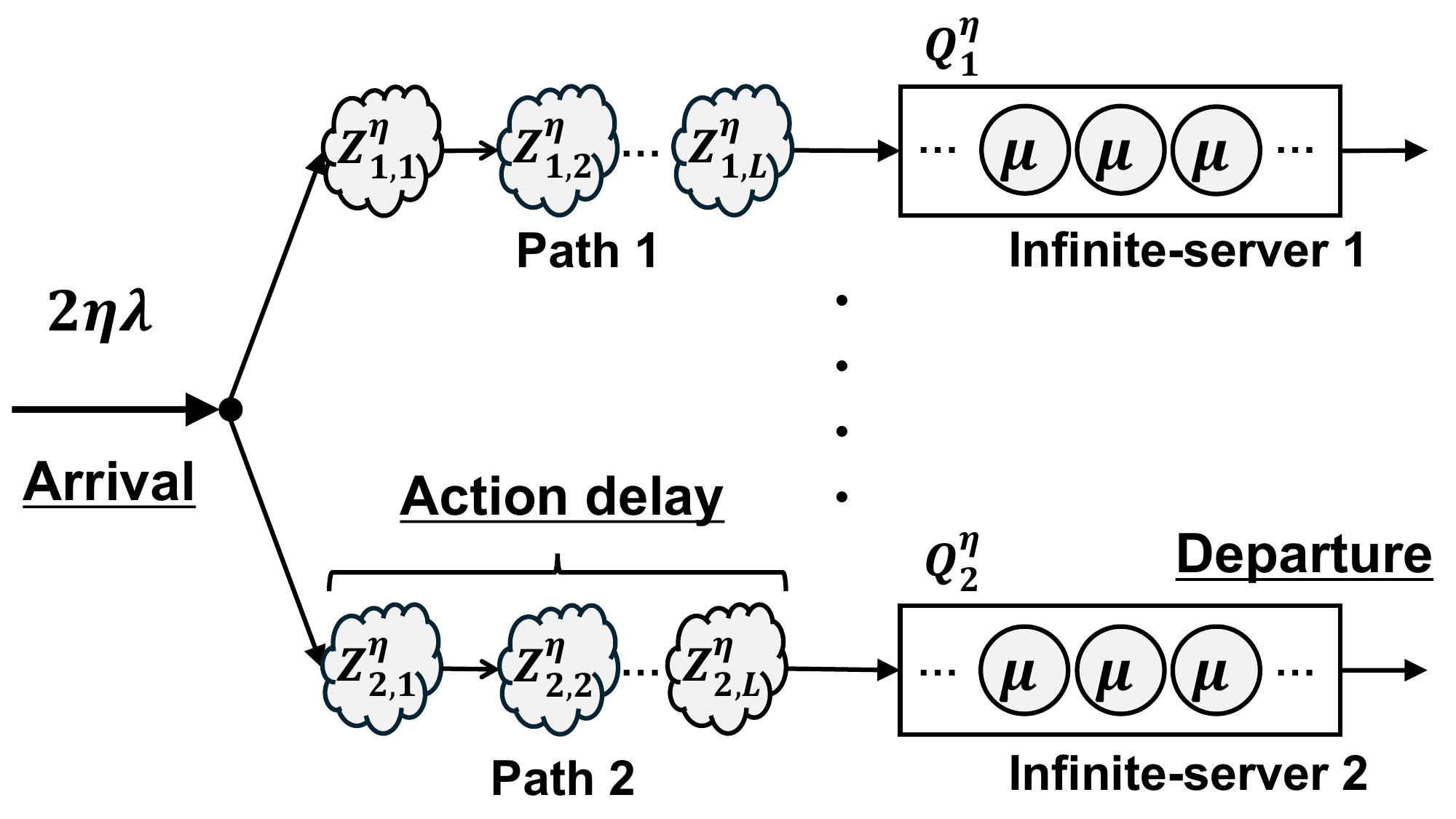}
    \caption{Overview of the system of two parallel infinite-server queues with action delay.}
    \label{fig:parallel-infinite-server}
\end{figure}

We describe a system of two parallel infinite-server queues with action delay as shown in Fig.~\ref{fig:parallel-infinite-server}. Service times at each infinite-server queue are exponentially distributed with rate $\mu>0$. Jobs arrive at a dispatcher according to a Poisson process with rate $2 \eta \lambda>0$, where $\lambda>0$ is fixed, and $\eta$ is a scaling parameter. Let $Q_i^\eta(t)$ denote the number of jobs in service at server $i$ at time $t$, and define the fluid scaled number of jobs by $\widehat{Q}_i^\eta(t):=Q_i^\eta(t)/\eta$. The purpose of load balancing here is to avoid uneven use of service resources by reducing the difference in the numbers of jobs in service between the two servers.

When a job arrives at the dispatcher, it is assigned to one of the two infinite-server queues based on the current scaled numbers of jobs in service. Under the multinomial-logit routing rule, the probability of selecting server $i$ is given by
\begin{equation}
p_i(\widehat{Q}_1^\eta(t), \widehat{Q}_2^\eta(t)) = \frac{\exp(-\theta \widehat{Q}_i^\eta(t))}{\exp(-\theta \widehat{Q}_1^\eta(t))+\exp(-\theta \widehat{Q}_2^\eta(t))},
\label{eq:multi-logit}
\end{equation}
where $\theta\geq 0$ is the sensitivity parameter. When $\theta=0$, a job is allocated uniformly at random, and as $\theta\to\infty$, \eqref{eq:multi-logit} smoothly approximates the join-the-shortest-queue (JSQ) policy. 

After being assigned, a job does not immediately enter service at the selected server. It enters service only after an action delay. For flexibility in the delay distribution, we assume that the action delay follows an Erlang distribution with $L \in \mathbb{N}$ phases and mean $D>0$. The transition rate of each phase is $\gamma=L/D$. A delayed job moves through the phase chain to reach the selected server. 
To track the number of jobs in the delay phases in the following section, let $Z_{i,\ell}^\eta(t)$ denote the number of jobs that have been routed to server $i$ and are in phase $\ell$ of their action delay at time $t$, and define $\widehat{Z}_{i,\ell}^\eta(t)$ by $\widehat{Z}_{i,\ell}^\eta(t):=Z_{i,\ell}^\eta(t)/\eta$.

\section{Fluid limit}
\label{sec:fluid}

In this section, we derive the fluid limit of the model described in Section~\ref{sec:model}. Since the jobs in the delay phases are retained as state variables, the scaled process forms a density-dependent Markov chain, and its law of large numbers yields a finite system of $2(L+1)$ ODEs that explicitly tracks the jobs in each delay phase (Subsection~\ref{subsec:density}). Then, we compute the symmetric equilibrium of this system (Subsection~\ref{subsec:equilibrium}), which serves as the reference point for the stability analysis in Section~\ref{sec:linearized_imbalance}.

\subsection{Density-dependent Markov chain and fluid limit}
\label{subsec:density}
We define the scaled state process by
\begin{equation*}
\widehat S^\eta(t)=\frac{1}{\eta}\bigl(Q_1^\eta(t),Q_2^\eta(t),Z_{1,1}^\eta(t),\ldots,Z_{1,L}^\eta(t), Z_{2,1}^\eta(t),\ldots, Z_{2,L}^\eta(t)\bigr), \qquad \eta\in\mathbb N.
\end{equation*}
For each possible jump direction $\nu$, the process $\widehat S^\eta$ jumps from a state $s$ to $s+\nu/\eta$ at rate $\eta\beta_\nu(s)$, where the functions $\beta_\nu$ are independent of $\eta$. The service-completion and delay-phase transition rates are proportional to the corresponding job counts, each of which is $\eta$ times its scaled value. The rate at which jobs are assigned to infinite-server queue $i$ is $2\eta\lambda$ times a routing probability that depends only on the scaled number of jobs in service. Each transition rate has the form $\eta\beta_\nu(s)$. Thus, $\{\widehat S^\eta\}_{\eta\in\mathbb N}$ is a
density-dependent family in the sense of
Ethier and Kurtz~\cite[Chapter~11]{ethier1986markov}.

The set of jump directions is finite, and the associated drift
\begin{equation*}
    \mathcal{F}(s)=\sum_\nu \nu\beta_\nu(s)
\end{equation*}
is globally Lipschitz. We construct the processes $\widehat S^\eta$ on a common probability space using independent unit-rate Poisson processes, one for each jump direction, shared across all $\eta$. Therefore, whenever $\widehat S^\eta(0)$ converges almost surely, Theorem~11.2.1 in Ethier and Kurtz~\cite{ethier1986markov} implies that $\widehat S^\eta$ converges almost surely, uniformly on compact time intervals, to the unique solution $s(t)$ of
\begin{equation*}
\dot s(t)=\mathcal{F}(s(t)), \quad s(0)=s_0.
\end{equation*}
Writing
\[
s(t)  =\bigl( q_1(t),q_2(t), z_{1,1}(t),\ldots,z_{2,L}(t)\bigr),
\]
the limiting equations are
\begin{align}
\dot z_{i,1}(t) &= 2\lambda p_i(q_1(t),q_2(t))-\gamma z_{i,1}(t), \label{eq:direct_first_phase} \\
\dot z_{i,\ell}(t) &= \gamma z_{i,\ell-1}(t)-\gamma z_{i,\ell}(t), \qquad \ell=2,\ldots,L, \label{eq:direct_middle_phase} \\
\dot q_i(t) &= \gamma z_{i,L}(t)-\mu q_i(t), \label{eq:direct_q}
\end{align}
for $i=1,2$. Here $q_i(t)$ is the fluid limit of the scaled number of jobs in service at server $i$, and $z_{i,\ell}(t)$ is that of the scaled number of jobs assigned to server $i$ that are in delay phase $\ell$. Equations \eqref{eq:direct_first_phase}--\eqref{eq:direct_q} describe the inflow of newly assigned jobs into the first delay phase and the outflow from that phase, the transitions between consecutive delay phases, and the entry of jobs into service after the last delay phase, together with the service completion at an infinite-server queue, respectively.

\begin{remark}
The fluid limit in~\cite{pender2020stochastic} is obtained for a model in which the routing probabilities depend on the number of jobs at a fixed amount of time earlier. The occupancy vector alone is generally not Markovian because its current value does not determine the routing rates. A history-valued state can restore the Markov property. The limit is a delay differential equation, so the proof relies on strong approximations for Poisson processes~\cite{kurtz1978strong}. The phase representation used here instead retains the jobs in the delay phases as state variables, so $\widehat{S}^\eta$ is a finite-dimensional Markov jump process, and the law of large numbers for density-dependent families applies directly.
\end{remark}

\subsection{Symmetric equilibrium}
\label{subsec:equilibrium}
We focus on the symmetric equilibrium solution. At equilibrium,
\begin{equation*}
q_1^*=q_2^*=q^*, \qquad z_{i,\ell}^*=z^*, \qquad i=1,2,\quad \ell=1,\ldots,L.
\end{equation*}
By symmetry, we have
\begin{equation*}
    p_1(q^*,q^*)=p_2(q^*,q^*)=\frac{1}{2}.
\end{equation*}
Therefore, each delay phase receives jobs at rate $\lambda$. Rearranging equations~\eqref{eq:direct_first_phase} and \eqref{eq:direct_middle_phase} gives the following equilibrium conditions for the delay phases:
\begin{align*}
    0=&\lambda-\gamma z_{i,1}^*, \\
    0=&\gamma z_{i,\ell-1}^*-\gamma z_{i,\ell}^*,
    \qquad \ell=2,\ldots,L.
\end{align*}
Thus, all delay phases have the same equilibrium value
\begin{equation*}
z_{i,\ell}^* = \frac{\lambda}{\gamma} = \frac{\lambda D}{L}, \qquad \ell=1,\ldots,L.
\label{eq:z_equilibrium}
\end{equation*}
Similarly, \eqref{eq:direct_q} gives
\begin{equation*}
    0=\gamma z_{i,L}^*-\mu q_i^*.
\end{equation*}
Since $\gamma z_{i,L}^*=\lambda$, we obtain the symmetric equilibrium
\begin{equation*}
q^* = \frac{\lambda}{\mu}. \label{eq:q_equilibrium}
\end{equation*}

\section{Linearized imbalance dynamics}
\label{sec:linearized_imbalance}
In this section, we analyze the imbalance between the two parallel infinite-server queues around the symmetric equilibrium obtained in Section~\ref{sec:fluid}. Exploiting the symmetry of the system, we first reduce the fluid dynamics to a closed system for the differences in the scaled numbers of jobs between the two parallel infinite-server queues, both in service and in each delay phase, and linearize it around the equilibrium (Subsection~\ref{subsec:difference_mode}). Then, we derive the characteristic equation for an arbitrary number of phases $L$ and show that it coincides with that of the Erlang delayed-information model (Subsection~\ref{subsec:general_L_characteristic}). We then examine exponential delays (Subsection~\ref{subsec:L1}), two-phase Erlang delays (Subsection~\ref{subsec:L2}), and the deterministic-delay limit (Subsection~\ref{subsec:deterministic_delay_limit}).

\subsection{Difference mode decomposition}
\label{subsec:difference_mode}

We consider the difference of the scaled numbers of jobs in service in the two parallel infinite-server queues and define
\begin{equation*}
x(t)=q_1(t)-q_2(t), \label{eq:x_imbalance}
\end{equation*}
and, for each delay phase,
\begin{equation*}
    w_\ell(t)=z_{1,\ell}(t)-z_{2,\ell}(t),
    \qquad \ell=1,\ldots,L.
    \label{eq:w_imbalance}
\end{equation*}
The variable $x(t)$ measures the imbalance in the numbers of jobs in service between the two parallel infinite-server queues, while $w_\ell(t)$ measures the corresponding imbalance in delay phase $\ell$. In what follows, we drop the argument $t$ where the meaning is clear.

The multinomial-logit rule \eqref{eq:multi-logit} satisfies the exact identity
\begin{equation}
p_1(q_1,q_2)-p_2(q_1,q_2) =-\tanh\!\left(\frac{\theta x}{2}\right), \label{eq:logit_tanh}
\end{equation}
which follows by multiplying the numerator and the denominator of
\eqref{eq:multi-logit} by $\exp\{\theta(q_1+q_2)/2\}$. Subtracting equations \eqref{eq:direct_first_phase}--\eqref{eq:direct_q} for $i=2$ from those for $i=1$ and using \eqref{eq:logit_tanh}, we obtain the nonlinear difference system
\begin{align}
\dot x(t) &=\gamma w_L(t)-\mu x(t), \label{eq:diff_x}\\
\dot w_1(t) &= -2\lambda\tanh\!\left(\frac{\theta x(t)}{2}\right)-\gamma w_1(t), \label{eq:diff_w1}  \\
\dot w_\ell(t) &= \gamma w_{\ell-1}(t)-\gamma w_\ell(t),\qquad \ell=2,\ldots,L. \label{eq:diff_wl}
\end{align}
The system \eqref{eq:diff_x}--\eqref{eq:diff_wl} is closed in the difference variables. Moreover, since $p_1+p_2=1$, the sum mode $q_1(t)+q_2(t)$ evolves independently of $\theta$. The linear system for deviations of the sum variables from equilibrium is triangular, with diagonal entries $-\gamma$ for the delay phases and $-\mu$ for the jobs in service, so the sum mode is asymptotically stable. Hence, the routing sensitivity acts only through the difference mode, and the imbalance dynamics can be studied in isolation.

The symmetric equilibrium of Section~\ref{sec:fluid} corresponds to $x^*=0$ and $w_\ell^*=0$ for all $\ell$. Since $\tanh(\theta x/2)=\theta x/2 +O(x^3)$, linearizing \eqref{eq:diff_x}--\eqref{eq:diff_wl} around this equilibrium gives
\begin{align}
\dot x(t) &=  \gamma w_L(t)-\mu x(t),\label{eq:lin_x}\\
\dot w_1(t) &= -\lambda\theta x(t)-\gamma w_1(t), \label{eq:lin_w1} \\
\dot w_\ell(t) &= \gamma w_{\ell-1}(t)-\gamma w_\ell(t), \qquad \ell=2,\ldots,L. \label{eq:lin_wl}
\end{align}

\begin{remark}
For $N$ homogeneous servers with total arrival rate $N \eta \lambda$, the linearization at the symmetric equilibrium separates into one average mode and $N-1$ identical imbalance modes. Each imbalance mode satisfies the same linear system \eqref{eq:lin_x}--\eqref{eq:lin_wl} as in the case of two parallel infinite-server queues. Hence, the characteristic equation \eqref{eq:char_polynomial} and the associated local stability conditions remain valid for any $N$, as in the delayed-information model~\cite{novitzky2020probabilistic}. However, the nonlinear reduction relies on the identity for the two parallel infinite-server queues~\eqref{eq:logit_tanh}. For $N>2$, a pairwise difference in routing probabilities also depends on the other servers, so the dynamics of the corresponding pairwise difference variables are generally not closed.
\end{remark}

In vector form, we introduce
\begin{equation*}
    y(t)=(x(t),w_1(t),\ldots,w_L(t))^\top.
\end{equation*}
Then, we have
\begin{equation}
    \dot y(t)=A_L y(t),
    \label{eq:AL_system}
\end{equation}
where
\begin{equation*}
A_L =
    \begin{pmatrix}
        -\mu & 0 & 0 & \cdots & 0 & \gamma \\
        -\lambda\theta & -\gamma & 0 & \cdots & 0 & 0 \\
        0 & \gamma & -\gamma & \cdots & 0 & 0 \\
        \vdots & & \ddots & \ddots & \vdots & \vdots \\
        0 & \cdots & 0 & \gamma & -\gamma & 0 \\
        0 & \cdots & 0 & 0 & \gamma & -\gamma
    \end{pmatrix}.
    \label{eq:AL_matrix}
\end{equation*}
Here, the entry $\gamma$ in the first row lies in the column corresponding to $w_L$, and represents the delayed feedback from the last delay phase into service.

\subsection{Characteristic equation for general $L$}
\label{subsec:general_L_characteristic}

We derive the characteristic equation of the $(L+1)$-dimensional linearized system \eqref{eq:AL_system}. We consider an eigenmode
\begin{equation*}
    y(t)=v\exp(rt), \qquad v=(X,W_1,\ldots,W_L)^\top\neq 0,
\end{equation*}
where $r\in\mathbb C$ is the spectral parameter. Substituting this form into
\eqref{eq:lin_x}--\eqref{eq:lin_wl} yields
\begin{align}
    (r+\mu)X &= \gamma W_L,  \label{eq:eig_x} \\
    (r+\gamma)W_1  &=  -\lambda\theta X,\label{eq:eig_w1} \\
    (r+\gamma)W_\ell &= \gamma W_{\ell-1}, \qquad \ell=2,\ldots,L. \label{eq:eig_wl}
\end{align}
We first assume $\theta>0$, in which case $r=-\gamma$ is not an eigenvalue. Applying the recursion \eqref{eq:eig_wl} to \eqref{eq:eig_w1} gives
\begin{equation}
    W_\ell  =  -\frac{\lambda\theta\gamma^{\ell-1}}{(r+\gamma)^\ell}X,  \qquad \ell=1,\ldots,L, \label{eq:Wl_general}
\end{equation}
and substituting the expression for $W_L$ into \eqref{eq:eig_x} yields
\begin{equation*}
(r+\mu)X = -\lambda\theta  \left(\frac{\gamma}{r+\gamma}\right)^L X.
\end{equation*}
For a nontrivial eigenmode, $X\neq 0$, and therefore
\begin{equation}
r+\mu +\lambda\theta\left(\frac{\gamma}{r+\gamma}\right)^L= 0,  \label{eq:characteristic_L}
\end{equation}
or equivalently, in polynomial form,
\begin{equation}
    (r+\mu)(r+\gamma)^L+\lambda\theta\gamma^L=0. \label{eq:char_polynomial}
\end{equation}
When $\theta=0$, the linearized system has a block-triangular structure, and the characteristic equation is
\begin{equation*}
(r+\mu)(r+\gamma)^L=0.
\end{equation*}
Thus, the polynomial characteristic equation~\eqref{eq:char_polynomial} remains valid for all $\theta\geq0$. Although the action-delay model considered here differs from the delayed-information model~\cite{novitzky2020probabilistic}, \eqref{eq:characteristic_L} coincides with the expression obtained for $N=2$ when the delay follows an Erlang distribution, after normalizing the arrival rate. Equation~\eqref{eq:Wl_general} also shows how the delay phases shape the feedback from routing decisions to the imbalance in the numbers of jobs in service between the two parallel infinite-server queues. For $\theta>0$, with the normalization $X=1$, the eigenvector associated with a root $r$ of \eqref{eq:char_polynomial} is
\begin{equation*}
    v(r) =  \left(1,\, -\frac{\lambda\theta}{r+\gamma},\,-\frac{\lambda\theta\gamma}{(r+\gamma)^2},\, \ldots,\,-\frac{\lambda\theta\gamma^{L-1}}{(r+\gamma)^L}\right)^\top. \label{eq:eigenvector_general}
\end{equation*}
Each factor $\gamma/(r+\gamma)$ is the response of a single exponential delay phase to the mode $\exp(rt)$, so the product $\gamma^L/(r+\gamma)^L$ in \eqref{eq:characteristic_L} encodes the attenuation and the phase shift induced by the whole $L$-phase chain. Thus, the stability of the imbalance dynamics depends on how the routing feedback is filtered through the delay phases.

\subsection{Exponential delay ($L=1$)}
\label{subsec:L1}

We first consider the case $L=1$, which corresponds to an exponentially distributed action delay with mean $D$, so $\gamma=1/D$. In this case, \eqref{eq:char_polynomial} reduces to
$(r+\mu)(r+\gamma)+\lambda\theta\gamma=0$, that is,
\begin{equation}
    r^2+(\mu+\gamma)r+\gamma(\mu+\lambda\theta)=0, \label{eq:L1_quadratic}
\end{equation}
whose roots are
\begin{equation*}
r_{\pm} =-\frac{\mu+\gamma}{2}\pm\frac12\sqrt{(\mu-\gamma)^2-4\gamma\lambda\theta}. \label{eq:eigenvalues_L1}
\end{equation*}
The trace of $A_1$ is $-(\mu+\gamma)<0$ and its determinant is
$\gamma(\mu+\lambda\theta)>0$ for all $\mu,\gamma>0$ and $\lambda\theta\ge 0$.
Hence, both roots have negative real parts, and the symmetric equilibrium is locally asymptotically stable for every choice of positive parameters. That is, an exponentially distributed action delay cannot destabilize the system.

The roots form a complex-conjugate pair if and only if
\begin{equation}
4\gamma\lambda\theta>(\mu-\gamma)^2, \label{eq:L1_complex_condition}
\end{equation}
in which case $r_{\pm}=-(\mu+\gamma)/2\pm i\omega$ with
\begin{equation*}
 \omega = \frac{1}{2} \sqrt{4\gamma\lambda\theta-(\mu-\gamma)^2}.
\end{equation*}
For the linearized system with initial conditions $x(0)=x_0$ and $w_1(0)=w_0$, the imbalance is then
\begin{equation*}
x(t) = \exp\left(-\frac{(\mu+\gamma)t}{2}\right) \left[ x_0\cos(\omega t) + \frac{ \gamma w_0+\frac{\gamma-\mu}{2}x_0}{\omega} \sin(\omega t)\right].
\label{eq:L1_damped_oscillation}
\end{equation*}
Thus, the imbalance decays at rate $\frac{\mu+\gamma}{2}$ while oscillating with period $\frac{2\pi}{\omega}$. An exponential action delay can produce damped oscillations of the imbalance in the number of jobs in service, but never sustained ones.

\subsection{Erlang delay ($L=2$)}
\label{subsec:L2}

When $L=2$, the characteristic equation is given by
\begin{equation*}
(r+\mu)(r+\gamma)^2+\lambda\theta\gamma^2=0. \label{eq:char_L2}
\end{equation*}
Equivalently,
\begin{equation}
r^3 + (\mu+2\gamma)r^2 + (2\mu\gamma+\gamma^2)r + \gamma^2(\mu+\lambda\theta) =0. \label{eq:cubic_L2}
\end{equation}
We write this cubic equation as
\begin{equation*}
    r^3+a_1r^2+a_2r+a_3=0,
\end{equation*}
where
\begin{equation*}
    a_1=\mu+2\gamma, \qquad  a_2=2\mu\gamma+\gamma^2, \qquad a_3=\gamma^2(\mu+\lambda\theta).
\end{equation*}
The Routh--Hurwitz stability criterion for a cubic polynomial states that all roots have negative real parts if and only if
\begin{equation*}
a_1>0,\quad a_2>0,\quad a_3>0, \text{ and } a_1a_2>a_3.
\end{equation*}
The first three inequalities hold for all positive parameters. Therefore, stability is determined by the condition
\begin{equation*}
(\mu+2\gamma)(2\mu\gamma+\gamma^2) > \gamma^2(\mu+\lambda\theta).
\end{equation*}
Equivalently,
\begin{equation}
    \lambda\theta<\frac{2\mu^2}{\gamma}+4\mu+2\gamma. \label{eq:L2_stability_condition}
\end{equation}
Thus, the two-phase linearized difference system is unstable if
\begin{equation}
    \lambda\theta >  \frac{2\mu^2}{\gamma}+4\mu+2\gamma. \label{eq:L2_instability_condition}
\end{equation}
At equality, the dominant roots lie on the imaginary axis, corresponding to the local stability boundary of the linearized system. Unlike $L=1$, a two-phase Erlang-distributed delay ($L=2$) can destabilize the symmetric equilibrium.

Condition \eqref{eq:L2_instability_condition} characterizes local instability of the symmetric fluid equilibrium. However, it does not characterize the stationary behavior of the system for finite $\eta$.

\subsection{Deterministic-delay limit ($L\to\infty$)}
\label{subsec:deterministic_delay_limit}

Finally, we let the number of phases grow while keeping the mean action delay $D$ fixed. Substituting $\gamma=\frac{L}{D}$ into \eqref{eq:characteristic_L} gives
\begin{equation*}
r+\mu+\lambda\theta\left(\frac{1}{1+\frac{rD}{L}}\right)^L=0.
\end{equation*}
Since $(1+rD/L)^{-L}\to\exp(-rD)$ as $L\to\infty$, we obtain
\begin{equation*}
    r+\mu+\lambda\theta \exp(-rD)=0. \label{eq:deterministic_delay_characteristic}
\end{equation*}
This is the characteristic equation of the corresponding deterministic-delay model with delay $D$, as described in \cite{pender2017queues}. 

\section{Numerical experiments}
\label{sec:numerical}

In this section, we present numerical examples that illustrate the preceding analysis. The stochastic system is simulated as a continuous-time Markov jump process. We implement the simulation as a discrete-event model using SimPy, with exponentially distributed interarrival times, delay-phase durations, and service times at the corresponding rates.
Unless stated otherwise, we initialize the system at $Q_1^\eta(0)=0$, $Q_2^\eta(0)=\eta$, and $Z_{i,\ell}^\eta(0)=0$, corresponding to the scaled initial condition $q_1(0)=0$, $q_2(0)=1$, and $z_{i,\ell}(0)=0$. Each simulation curve shows a single sample path over $t\in[0, 20]$, and the fluid equations \eqref{eq:direct_first_phase}–\eqref{eq:direct_q} are integrated from the same initial condition using the explicit Euler method with a fixed time step of $5\times10^{-4}$.

\begin{figure}[H]
  \begin{subfigure}[b]{0.49\textwidth}
    \centering
    \includegraphics[width=\textwidth]{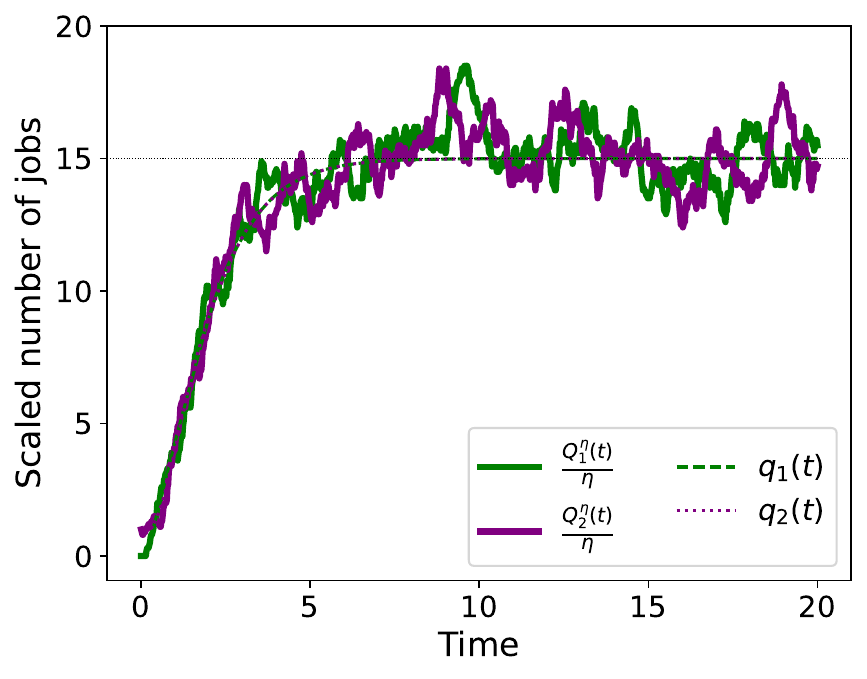}
    \caption{$\eta=10$.}\label{direct_eta10}
  \end{subfigure}
  \hfill
  \begin{subfigure}[b]{0.49\textwidth}
    \centering
    \includegraphics[width=\textwidth]{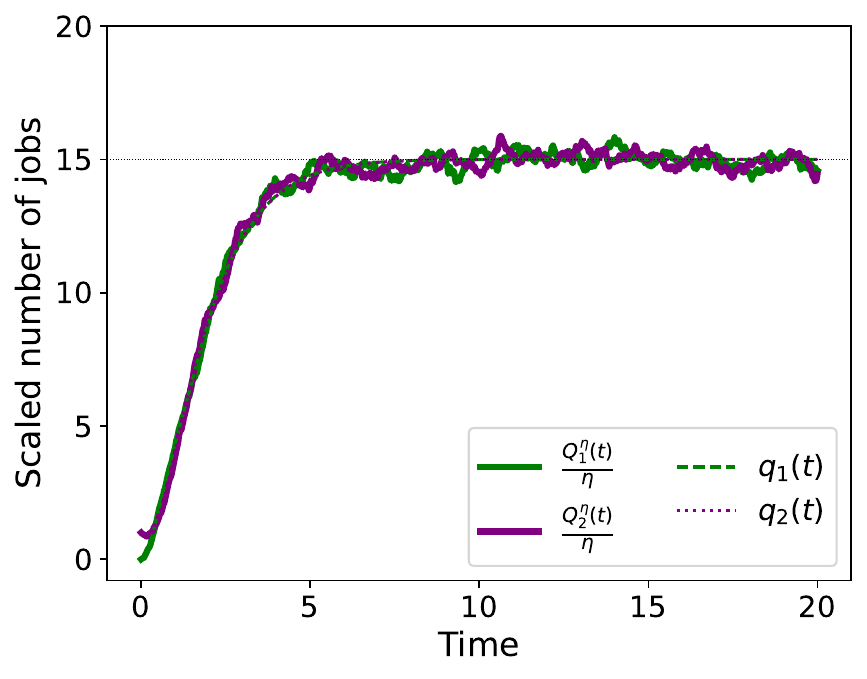}
    \caption{$\eta=100$.}\label{direct_eta100}
  \end{subfigure}
  \caption{Scaled numbers of jobs in service at each server for two values of the scaling parameter $\eta$ ($\lambda=15$, $\mu=1.0$, $D=1.0$, $\theta=1.0$, $L=1$).}\label{fig:direct_eta}
\end{figure}

Fig.~\ref{fig:direct_eta} compares the scaled numbers of jobs in service for two values of $\eta$ when the action delay follows an exponential distribution ($L=1$). For these parameter values, the linearized imbalance exhibits damped oscillations because the roots of \eqref{eq:L1_quadratic} form a complex-conjugate pair with negative real parts.
Moreover, the fluctuations around the fluid limit decrease as $\eta$ grows. When many jobs are present, the randomness of the individual arrivals, action delays, and service completions averages out, so the scaled number of jobs behaves almost deterministically. This averaging is, in effect, the law of large numbers for density-dependent Markov chains~\cite{ethier1986markov}, which is why the fluctuations shrink as the scale increases.

\begin{figure}[H]
\centering
\includegraphics[width=0.60\textwidth]{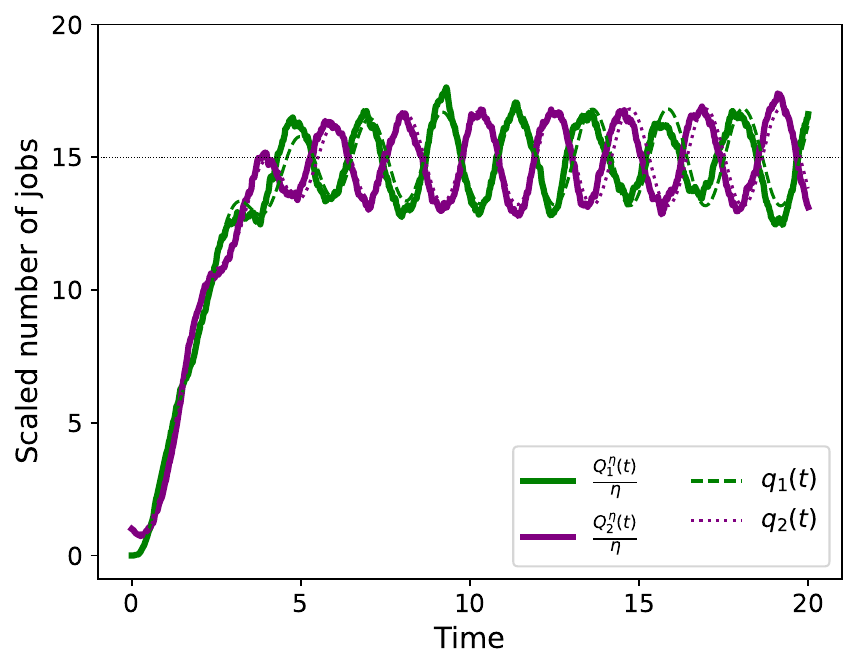}
  \caption{Scaled number of jobs in service at each server ($\lambda=15$, $\mu=1.0$, $D=1.0$, $\theta=1.0$, $L=2$, $\eta=100$).}\label{fig:direct}
\end{figure}

Fig.~\ref{fig:direct} compares the scaled numbers of jobs from stochastic simulation with the corresponding fluid trajectories for $L=2$. As in Fig.~\ref{fig:direct_eta}, the fluid limit agrees closely with the simulation results. The scaled number of jobs in service at each server oscillates around the equilibrium. The parameters satisfy the instability condition \eqref{eq:L2_instability_condition}, so the symmetric equilibrium is locally unstable.

\begin{figure}[H]
  \begin{subfigure}[b]{0.49\textwidth}
    \centering
    \includegraphics[width=\textwidth]{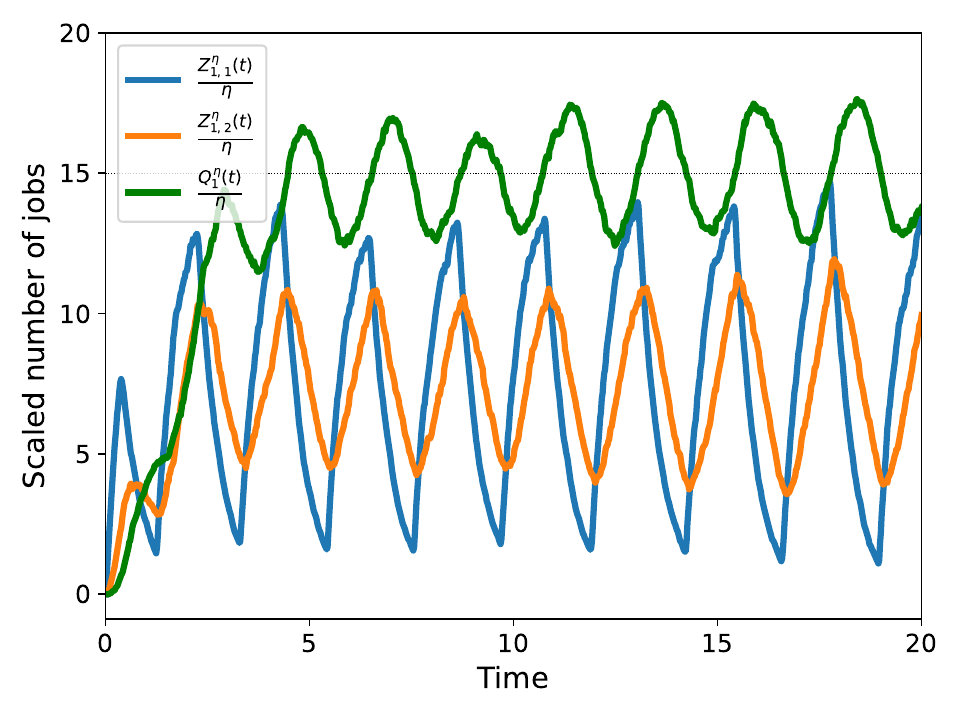}
    \caption{Stochastic simulation.}\label{direct_sim}
  \end{subfigure}
  \hfill
  \begin{subfigure}[b]{0.49\textwidth}
    \centering
    \includegraphics[width=\textwidth]{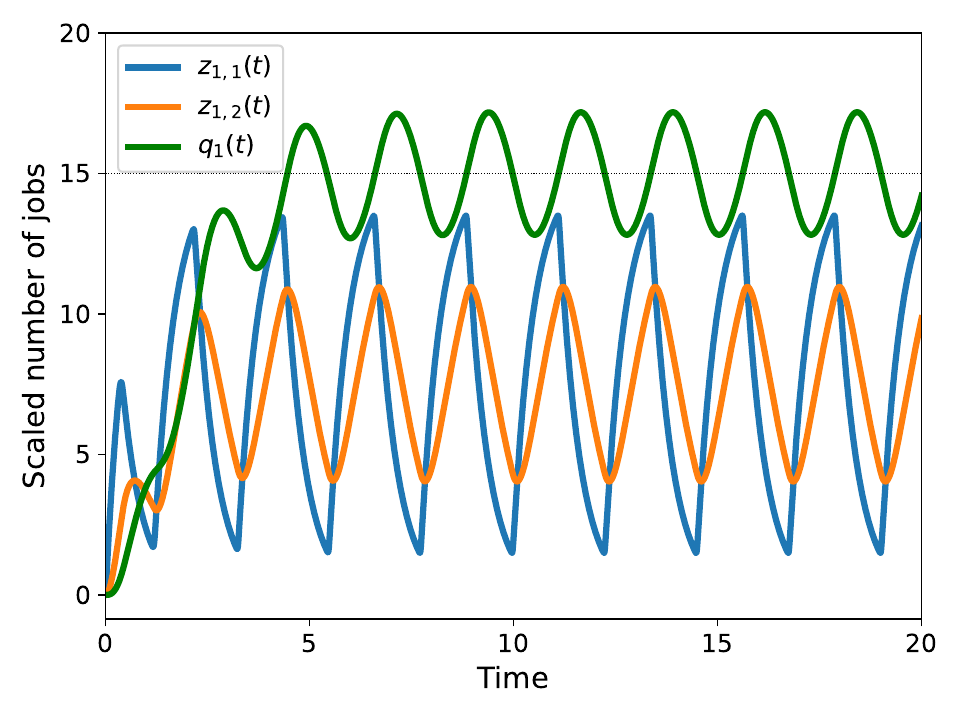}
    \caption{Fluid limit.}\label{direct_fluid}
  \end{subfigure}
  \caption{Scaled numbers of jobs in the delay phases and in service, obtained from the simulation and its fluid limit ($\lambda=15$, $\mu=1.0$, $D=1.0$, $\theta=1.0$, $\eta=100$).}\label{fig:direct_phase}
\end{figure}

Fig.~\ref{fig:direct_phase} depicts the scaled numbers of jobs in the delay phases and in service, obtained from the simulation and fluid limit for $L=2$. The parameter values are the same as in Fig.~\ref{fig:direct}. Because the phase representation retains the jobs in the delay phases as state variables, it makes the dynamics inside the delay visible. The main feature is that the oscillation propagates as a wave through the delay phases before it reaches the server. First, a routing decision disturbs the first delay phase. This disturbance then passes to the second phase and is subsequently reflected in the number of jobs in service. As a result, the peaks of $z_{1,1}$, $z_{1,2}$, and $q_1$ appear in this order, each later than the previous one. Each phase thus adds a small delay to the signal.

In this example, the additional lag before the oscillation appears in the number of jobs in service is larger than the lag between consecutive delay phases; the number of jobs in service responds at rate $\mu$, whereas the number of jobs in each delay phase responds at rate $\gamma$. This difference appears directly in the characteristic equation \eqref{eq:char_polynomial}, where the delay phases and the server enter through the factors $(r+\gamma)$ and $(r+\mu)$, respectively.

\begin{figure}[H]
  \begin{subfigure}[b]{0.49\textwidth}
    \centering
    \includegraphics[width=\linewidth]{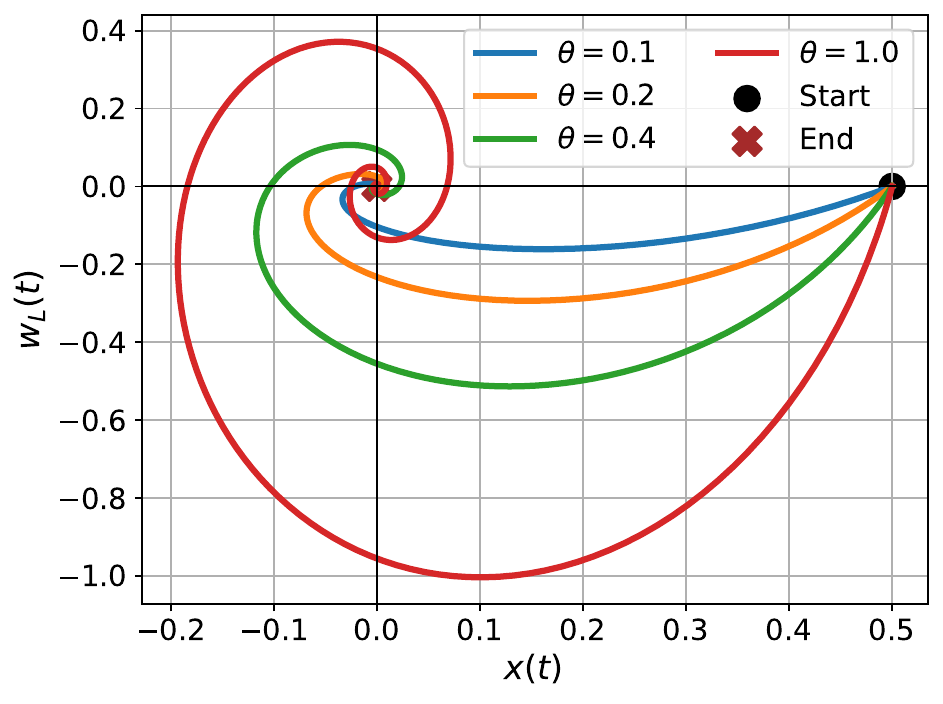}
    \caption{$\theta=0.1,0.2,0.4, 1.0$.}\label{phase0}
  \end{subfigure}
  \hfill
  \begin{subfigure}[b]{0.49\textwidth}
    \centering
    \includegraphics[width=\linewidth]{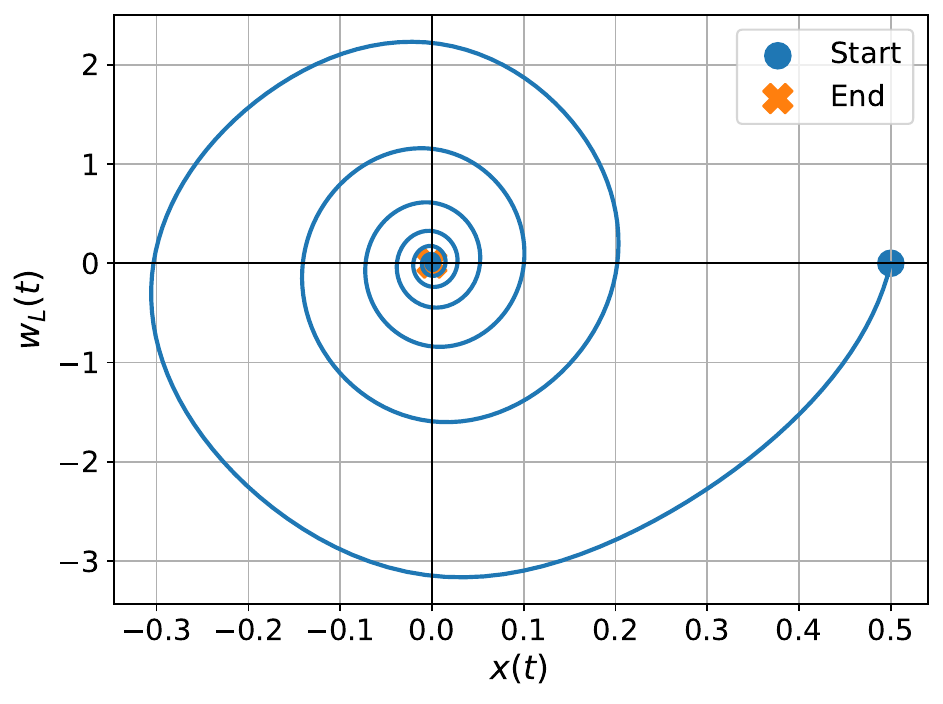}
    \caption{$\theta=10$.}\label{phase10}
  \end{subfigure}
  \caption{Phase plots of $(x(t), w_1(t))$ for varying sensitivity $\theta$, started at $(0.5,0)$ ($\lambda=10$, $\mu=1.0$, $D=1.0$, $L=1$).}\label{fig:phase_theta}
\end{figure}

Fig.~\ref{fig:phase_theta} shows the phase plots of $(x(t), w_1(t))$ for $L=1$ as the sensitivity parameter $\theta$ varies, with all trajectories started at $(0.5, 0)$. As $\theta \to 0$, the routing becomes independent of the number of jobs at each server. As $\theta$ grows, the policy approaches JSQ.
The left panel of Fig.~\ref{fig:phase_theta} shows the phase plots for small $\theta$. Each trajectory converges to the equilibrium, and the magnitude of the excursion in $w_1$ increases with $\theta$, since the delay phase imbalance $w_1(t)$ is only weakly coupled to the number of jobs in service when the routing is nearly independent of them.

The right panel of Fig.~\ref{fig:phase_theta} shows the case of a large sensitivity parameter ($\theta = 10$), for which the routing rule approximates the JSQ policy. Here, the roots form a complex-conjugate pair whose imaginary part is large compared with the decay rate $(\mu+\gamma)/2$, so the trajectory spirals into the equilibrium over several revolutions. The solution still converges, but only after many oscillations. In the linearized system, increasing $\theta$ raises the oscillation frequency while leaving the decay rate unchanged.

\begin{figure}[H]
  \begin{subfigure}[b]{0.49\textwidth}
    \centering
    \includegraphics[width=\textwidth]{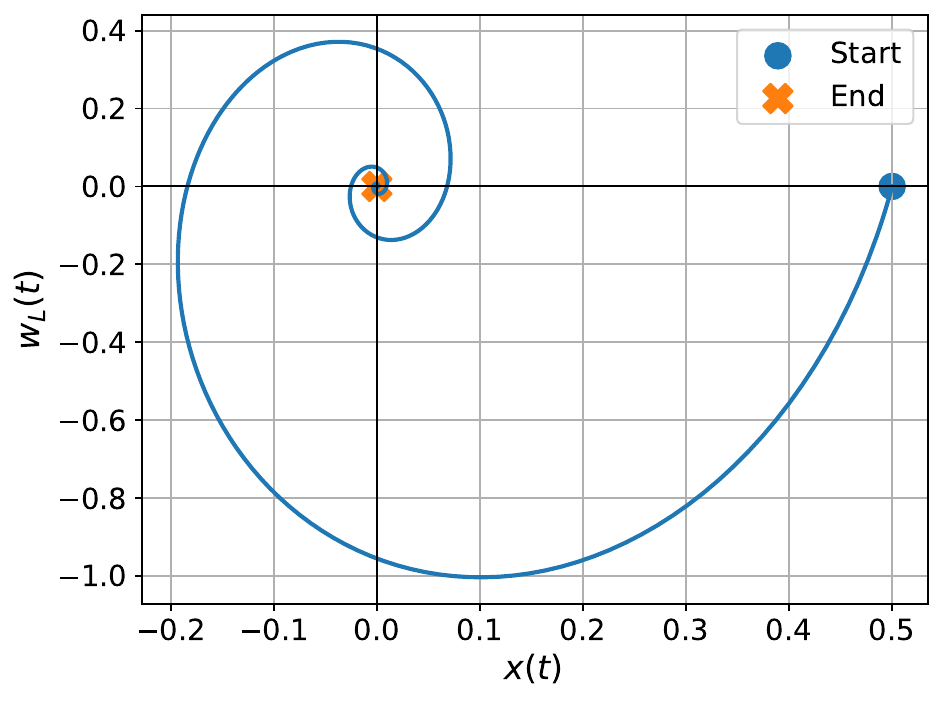}
    \caption{$L=1$.}\label{phase1}
  \end{subfigure}
  \hfill
  \begin{subfigure}[b]{0.49\textwidth}
    \centering
    \includegraphics[width=\textwidth]{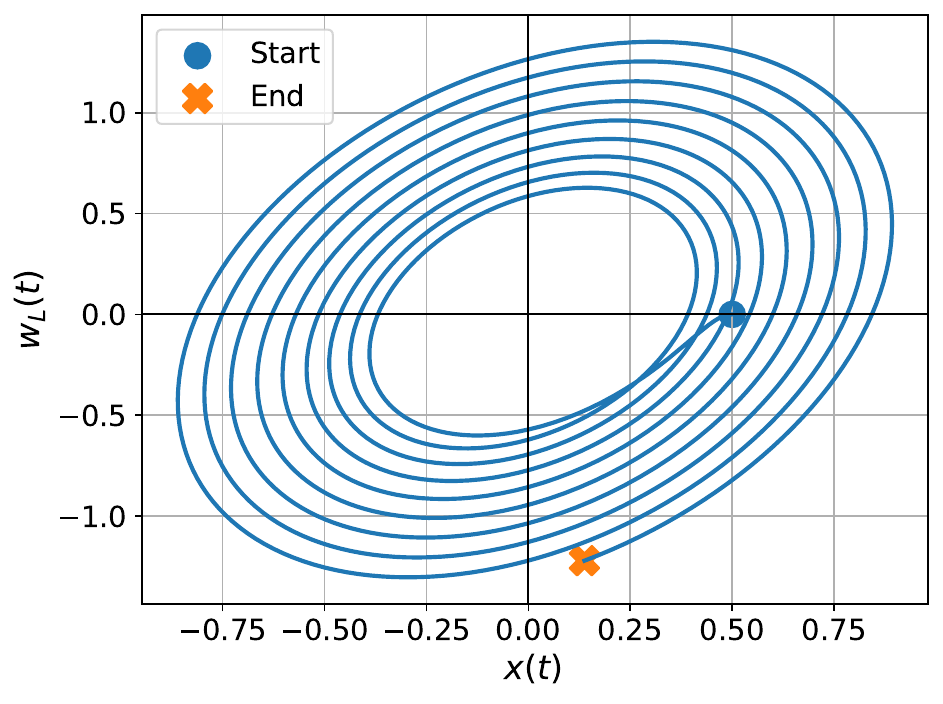}
    \caption{$L=2$.}\label{phase2}
  \end{subfigure}
    \begin{subfigure}[b]{0.49\textwidth}
    \centering
    \includegraphics[width=\textwidth]{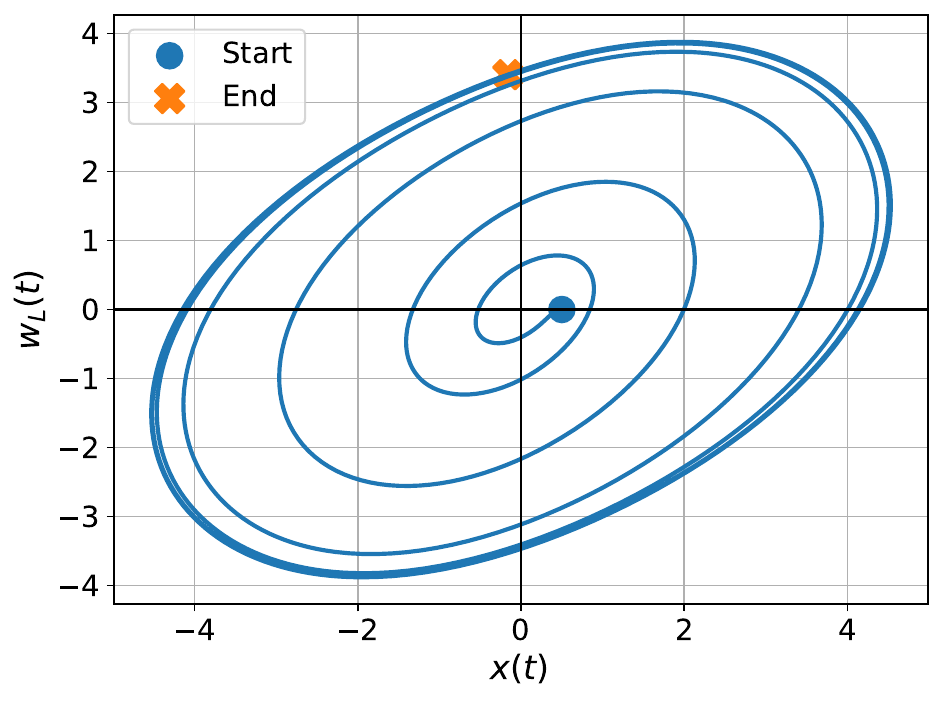}
    \caption{$L=3$.}\label{phase3}
  \end{subfigure}
  \hfill
  \begin{subfigure}[b]{0.49\textwidth}
    \centering
    \includegraphics[width=\textwidth]{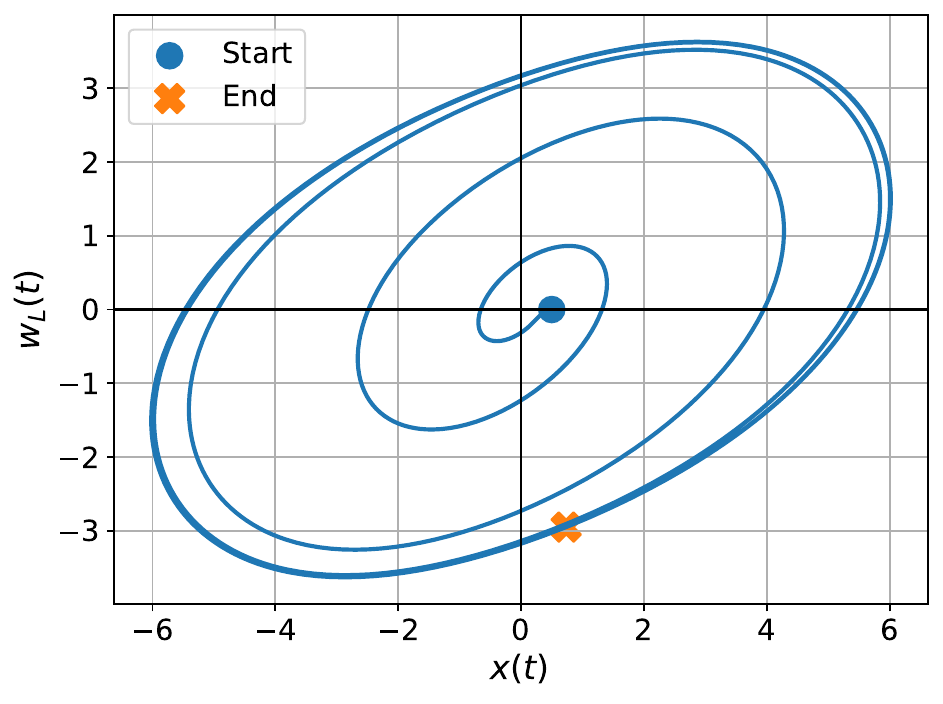}
    \caption{$L=4$.}\label{phase4}
  \end{subfigure}
  \caption{Phase plot of the imbalance $(x(t), w_L(t))$ for four values of the number of phases $L$, started at $x(0)=0.5$ and $w_\ell(0)=0$ for $\ell=1,\ldots,L$ and integrated over $t \in [0, 20]$ ($\lambda=10$, $\mu=1.0$, $D=1.0$, $\theta=1.0$).}\label{fig:phase_L}
\end{figure}

Fig.~\ref{fig:phase_L} shows the phase plots of $(x(t), w_L(t))$ for four values of the number of phases $L$. Each trajectory starts at the blue point $(0.5, 0)$, and the orange cross marks its position at $t=20$. The symmetric equilibrium is the origin. In all panels, the trajectory revolves around the equilibrium, reflecting the fact that the dominant roots of the characteristic equation form a complex-conjugate pair. For $L=1$, their real parts are negative, so the trajectory spirals inward and converges to the equilibrium, which is a stable focus. For $L=2$, the dominant pair has positive real parts, so the trajectory spirals outward and the equilibrium is unstable.

For $L=3$ and $L=4$ (the bottom two panels), the equilibrium is more strongly unstable. The projected trajectories suggest persistent oscillatory behavior over the simulated interval. As the imbalance grows, the routing feedback $2\lambda\tanh(\theta x/2)$ saturates, remaining bounded in absolute value by $2\lambda$. For the parameter values and time interval shown, the excursion in the displayed projection is larger for $L=4$ than for $L=3$.

\begin{figure}[H]
    \centering
    \includegraphics[width=0.75\linewidth]{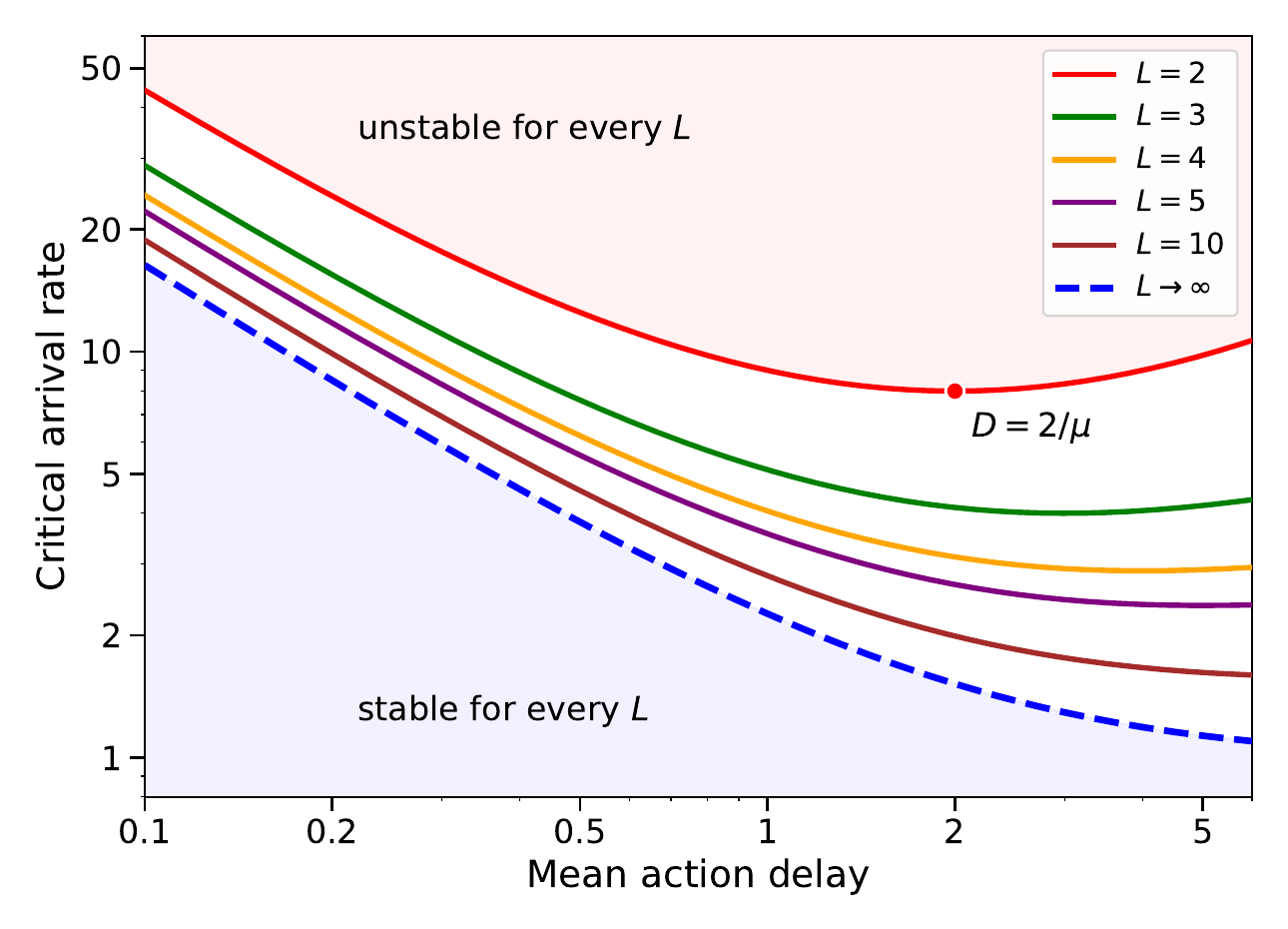}
    \caption{Local stability boundaries for different numbers of delay phases ($\mu=1.0$, $\theta=1.0$).}
    \label{fig:stability_region}
\end{figure}

Fig.~\ref{fig:stability_region} shows the critical arrival rate against the mean delay for several numbers of phases. For $L>2$, the stability boundaries are computed numerically from \eqref{eq:char_polynomial} by locating the smallest arrival rate at which the largest real part among its roots reaches zero. The region below each curve is locally stable. As $L$ increases, the stability region shown in the figure shrinks, and its boundary approaches the deterministic-delay boundary. There is no curve for $L=1$, because both eigenvalues of $A_1$ have negative real parts for every arrival rate.

From the Routh--Hurwitz stability criterion for a cubic polynomial in Subsection~\ref{subsec:L2}, the critical arrival rate for $L=2$ and $\theta>0$ is given by 
\begin{equation*}
    \lambda_c(D)=\frac{1}{\theta} \Bigl(\frac{4}{D} + 4 \mu + \mu^2 D \Bigr ),
\end{equation*}
which attains its minimum $8 \mu/\theta$ at $D=2/\mu$. For $\theta=0$, there is no finite critical arrival rate.

\begin{figure}[H]
    \centering
    \includegraphics[width=0.60\linewidth]{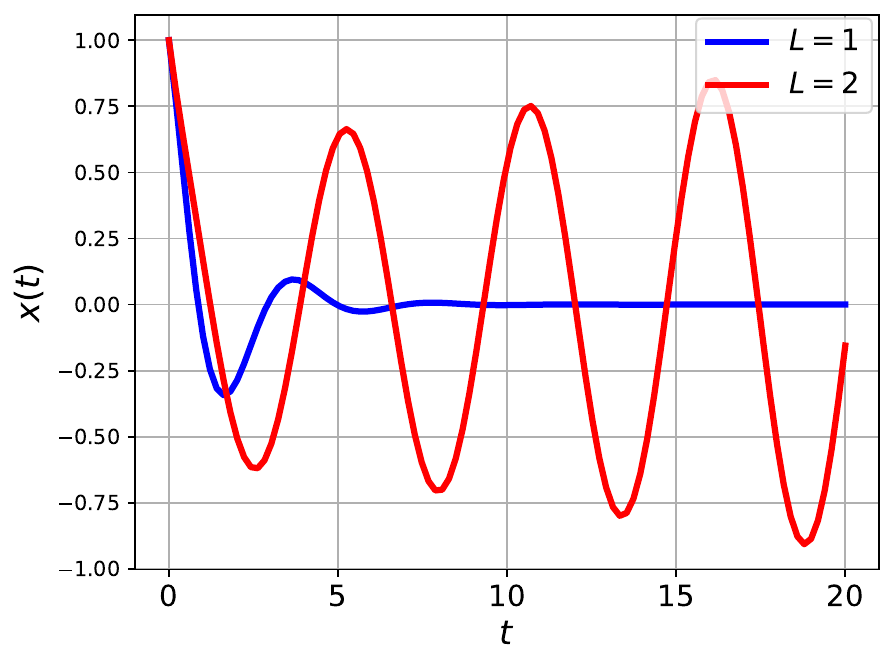}
    \caption{Imbalance $x(t)$ in the linearized system for $L=1$ and $L=2$ ($\lambda=10$, $\mu=1.0$, $D=4.0$, $\theta=1.0$, $x(0)=1.0$, $w_\ell(0)=0$).}\label{fig:diff}
\end{figure}

Fig.~\ref{fig:diff} shows the imbalance $x(t)$ obtained from the difference system \eqref{eq:diff_x}--\eqref{eq:diff_wl} for $L=1$ and $L=2$.
As shown in Section~\ref{subsec:L1}, for $L=1$ both roots of \eqref{eq:L1_quadratic} have negative real parts, so the imbalance decays to zero. For $L=2$, the stability condition \eqref{eq:L2_stability_condition} is violated, and the amplitude of the linearized oscillations grows. The linearized solution approximates the nonlinear solution while the imbalance remains sufficiently small. In the linearized system, the amplitude of the dominant mode grows at rate $\mathrm{Re}(r)$, where $r$ is a dominant root of \eqref{eq:cubic_L2}. The linearization $\tanh(\theta x/2)\approx \theta x/2$ is valid only for $|\theta x/2|\ll1$. As \(\lvert x(t)\rvert\) increases, the magnitude of the nonlinear routing feedback approaches $2\lambda$. The linearized solution grows over time, whereas the nonlinear imbalance remains bounded.

\section{Conclusion}
\label{sec:conclusion}

In this work, we analyzed parallel infinite-server queueing systems with an action delay, where the delay represents the lag between a routing decision and its execution. Representing the delay by an Erlang phase structure, we derived a finite-dimensional fluid limit that explicitly tracks the jobs in the delay phases. Then, we decomposed the dynamics into sum and difference modes and obtained the characteristic equation of the linearized imbalance dynamics for an arbitrary number of phases $L$. The phase-state formulation recovers the local stability boundary previously obtained for the Erlang delayed-information model~\cite{novitzky2020probabilistic}.
The numerical experiments support this analysis. First, the fluid trajectories closely follow the simulated sample paths over the time interval shown for both the stable and unstable parameter settings. Second, in the numerical examples with fixed mean delay, larger phase counts produced larger oscillations in the service imbalance.

These results suggest several directions for future work. First,  could evaluate load balance in terms of the rate of convergence to equilibrium, the tolerance within which the system is deemed balanced, or the stationary variance of the imbalance. Second, we could examine how heterogeneous service rates and heterogeneous delays across servers affect the equilibrium and the mode structure. Finally, we could investigate whether a routing rule that accounts for jobs in the delay phases can suppress oscillations. The appropriate criterion for such a rule, whether convergence rate, tolerance, or stationary variance of the imbalance, is an open question.

\bibliographystyle{splncs04}
\bibliography{ref}

@article{pender2020stochastic,
  title={A stochastic analysis of queues with customer choice and delayed information},
  author={Pender, Jamol and Rand, Richard and Wesson, Elizabeth},
  journal={Mathematics of Operations Research},
  volume={45},
  number={3},
  pages={1104--1126},
  year={2020},
  publisher={INFORMS}
}

@article{pender2017queues,
  title={Queues with choice via delay differential equations},
  author={Pender, Jamol and Rand, Richard H and Wesson, Elizabeth},
  journal={International Journal of Bifurcation and Chaos},
  volume={27},
  number={04},
  pages={1730016},
  year={2017},
  publisher={World Scientific}
}

@article{he2025randomized,
  title={Randomized Routing to Remote Queues},
  author={He, Shuangchi and Yang, Yunfang and Yu, Yao},
  journal={arXiv preprint arXiv:2505.04942},
  year={2025}
}

@article{novitzky2020limiting,
  title={Limiting the oscillations in queues with delayed information through a novel type of delay announcement},
  author={Novitzky, Sophia and Pender, Jamol and Rand, Richard H and Wesson, Elizabeth},
  journal={Queueing Systems},
  volume={95},
  number={3},
  pages={281--330},
  year={2020},
  publisher={Springer}
}

@article{mitzenmacher2000useful,
  title={How Useful Is Old Information?},
  author={Mitzenmacher, Michael},
  journal={IEEE Transactions on Parallel \& Distributed Systems},
  volume={11},
  number={01},
  pages={6--20},
  year={2000},
  publisher={IEEE Computer Society}
}

@article{abe2026mean,
  title={Mean-Field Analysis of Large-Scale Bipartite Queueing Models for Threshold-Based and Stochastic Offloading Multi-Access Edge Computing},
  author={Abe, Kazuma and Phung-Duc, Tuan},
  journal={Methodology and Computing in Applied Probability},
  volume={28},
  number={2},
  pages={57},
  year={2026},
  publisher={Springer}
}

@article{beraldi2022impact,
title = {On the impact of stale information on distributed online load balancing protocols for edge computing},
journal = {Computer Networks},
volume = {210},
pages = {108935},
year = {2022},
issn = {1389-1286},
author = {Roberto Beraldi and Claudia Canali and Riccardo Lancellotti and Gabriele Proietti Mattia}
}

@inproceedings{tahir2022learning,
  title={Learning mean-field control for delayed information load balancing in large queuing systems},
  author={Tahir, Anam and Cui, Kai and Koeppl, Heinz},
  booktitle={Proceedings of the 51st International Conference on Parallel Processing},
  pages={1--11},
  year={2022}
}

@book{ethier1986markov,
  title={Markov Processes: Characterization and Convergence},
  author={Ethier, Stewart N. and Kurtz, Thomas G.},
  publisher={Wiley},
  address={New York},
  year={1986}
}

@article{kurtz1978strong,
  title={Strong approximation theorems for density dependent {M}arkov chains},
  author={Kurtz, Thomas G},
  journal={Stochastic Processes and their Applications},
  volume={6},
  number={3},
  pages={223--240},
  year={1978},
  publisher={Elsevier}
}

@misc{novitzky2020probabilistic,
  title={Queues with delayed information: A probabilistic perspective},
  author={Novitzky, Sophia and Pender, Jamol},
  year={2020},
  note={Working paper}
}

@article{whitt2021many,
  title={On the many-server fluid limit for a service system with routing based on delayed information},
  author={Whitt, Ward},
  journal={Operations Research Letters},
  volume={49},
  number={3},
  pages={316--319},
  year={2021},
  publisher={Elsevier}
}
\end{document}